\documentclass[9pt,twocolumn,twoside]{opticajnl}

\journal{opticajournal} 

\setboolean{shortarticle}{true}

\title{Laser Mpemba effect}

\author[1,2,*]{Stefano Longhi}

\affil[1]{Dipartimento di Fisica, Politecnico di Milano, Piazza L. da Vinci 32, I-20133 Milano, Italy}
\affil[2]{IFISC (UIB-CSIC), Instituto de Fisica Interdisciplinar y Sistemas Complejos - Palma de Mallorca, Spain}

\affil[*]{stefano.longhi@polimi.it}

\begin{abstract}
This work explores the emergence of Mpemba-like effects within the quantum theory of lasers. By examining the temporal dynamics of photon number statistics in a single-mode laser above threshold, we reveal the curious and counterintuitive possibility that a laser system, starting with photon statistics far from equilibrium, may reach its stationary nearly-Poissonian distribution faster than a system initially closer to equilibrium. Drawing parallels to both classical and quantum Mpemba effects, we suggest that this behavior results from the unique relaxation dynamics of photon states, which is described by a non-integrable birth-death process. Our findings offer new insights into the foundational aspects of quantum laser light and contribute to the expanding body of research on non-equilibrium phenomena in quantum systems.
\end{abstract}

\setboolean{displaycopyright}{false} 

\begin{document}

\maketitle

{\em Introduction.} 
The quantum theory of the laser \cite{R1,R2} has long been a cornerstone of understanding light-matter interactions, providing profound insights into the statistical properties of laser light (see e.g. \cite{R3,R4,R5,Carmichael}). Unlike thermal light, which exhibits a Bose-Einstein distribution of photon statistics characterized by bunching behavior and large intensity fluctuations, laser light in its steady-state regime is defined by its coherence and nearly Poissonian photon number statistics \cite{R1,R3,R4,R5}. The phase transition from thermal to coherent light \cite{R5a,R5b,R5bis,R5c,R5d}  reflects the underlying quantum processes that govern the laser's operation from below to above threshold.
  A notorious quantum model of the laser is the Scully-Lamb model \cite{R1}, which is basically a birth-death master equation for photon population and coherences \cite{Carmichael}. Besides of providing a paradigmatic framework to study the statistical properties of laser light, this model
 is of major relevance in such diverse fields as Bose-Einstein condensation \cite{R6,R7,R8,R9}, black hole radiation \cite{R10,R11}, and in biological models of population birth-death processes \cite{bd}. 
The dynamical pathways through which a photon laser distribution from arbitrary initial conditions relaxes to the coherent, steady-state distribution has received little attention so far \cite{R12,R13,R14,R15} and remains an intriguing area of study. Typically, the transient buildup of laser light is assumed to begin from the vacuum or thermal state in the cavity mode \cite{R12}, and experimental techniques to accurately measure the transient statistical dynamics have been developed since the early days of laser research \cite{R5bis}. However, 
in a 'Gedanken' experiment or under quench dynamics one could consider other initial non-equilibrium photon field states in the cavity mode. The interest to investigate relaxation dynamics under general initial conditions goes beyond the quantum laser model, and could be of relevance to other systems, such as in  the non-equilibrium behavior of an 'atom laser' \cite{R6,R16}
exhibiting a truly laser-like behavior \cite{R17} and in the study of population dynamics in ecology and biological models \cite{bd}.\\
In recent years, the study of non-equilibrium statistical physics has uncovered counterintuitive phenomena, one of which is the Mpemba effect. Originally observed in the context of water freezing \cite{R18,R19} this effect refers to the paradoxical situation where a system initially farther from equilibrium reaches the steady state faster than one closer to it (see e.g. \cite{R20,R21} and references therein). Beyond its classical origins, there has been growing interest in exploring the quantum Mpemba effect, which involves anomalous relaxation behavior in quantum systems \cite{R22,R23,R24,R25,R26,R27,R28,R29,R30,R31,R32,R33,R34,R34b}. These works suggest that the quantum Mpemba effect can arise from unique features of quantum dynamics, such as interference, non-commuting observables, and quantum correlations, making it a fertile ground for theoretical and experimental exploration.
 Recent studies have highlighted the appearance of Mpemba-like phenomena and accelerated relaxation in photonic models \cite{R35,R36,R37}, such as in the decay process of quantum light in a passive optical cavity \cite{R36,R37} which is described by an exactly integrable birth-death markovian process \cite{R37}. Unlike such previous models, the Scully-Lamb laser model is not integrable \cite{R13} owing to nonlinearities arising from gain saturation. A natural question arises: can the Mpemba effect emerge in the quantum theory of the laser above threshold?\\ 
In this work, we investigate the dynamical evolution of photon number statistics in a single-mode laser within the framework of the Scully-Lambd model. We focus on the intriguing possibility that initial conditions far from equilibrium can result in a faster relaxation to the stationary nearly-Poissonian distribution compared to initial conditions closer to equilibrium.
These findings shed new light on the temporal behavior of lasers transitioning to steady-state operation at the quantum level, contributing to the growing body of work exploring non-equilibrium effects in quantum systems.\\
\\
{\it Scully-Lamb model and photon relaxation dynamics.}  The Scully-Lamb model describes the relaxation dynamics of the reduced density matrix  $\rho(t)$ of the photon field in a single-mode class-A laser in terms of a quantum master equation \cite{R1,R3,R4,R5bis,R6}. In the photon number (Fock) state basis $|n \rangle$, the density matrix elements $\rho_{n,n+r}(t)=\langle n | \rho (t) |n+r \rangle$ satisfy a set of equations which are decoupled in $r$ (see e.g. \cite{R4}, Chap.12) and take the form of a birth-death process \cite{Carmichael}, i.e. a continuous-time Markov process that is often used to study population dynamics in ecology, genetics, and biology \cite{bd}. For $r=0$, the equations describe the relaxation dynamics of the photon number distribution $P_n(t)=\rho_{n,n}(t)$ toward the stationary state $P_n^{(S)}$, while for $r>0$ they describe the damping of coherences (phase diffusion) responsible for the quantum limit of the laser linewidth.
Here we focus our attention to the dynamics of the photon number distribution, which is governed by the following birth-death master equation [Fig.1(a)]
\begin{equation}
\frac{dP_n}{dt}=G_nP_{n-1}+L_{n+1}P_{n+1}-(L_n+G_{n+1})P_n
\end{equation}
  \begin{figure}
\centering
    \includegraphics[width=0.48\textwidth]{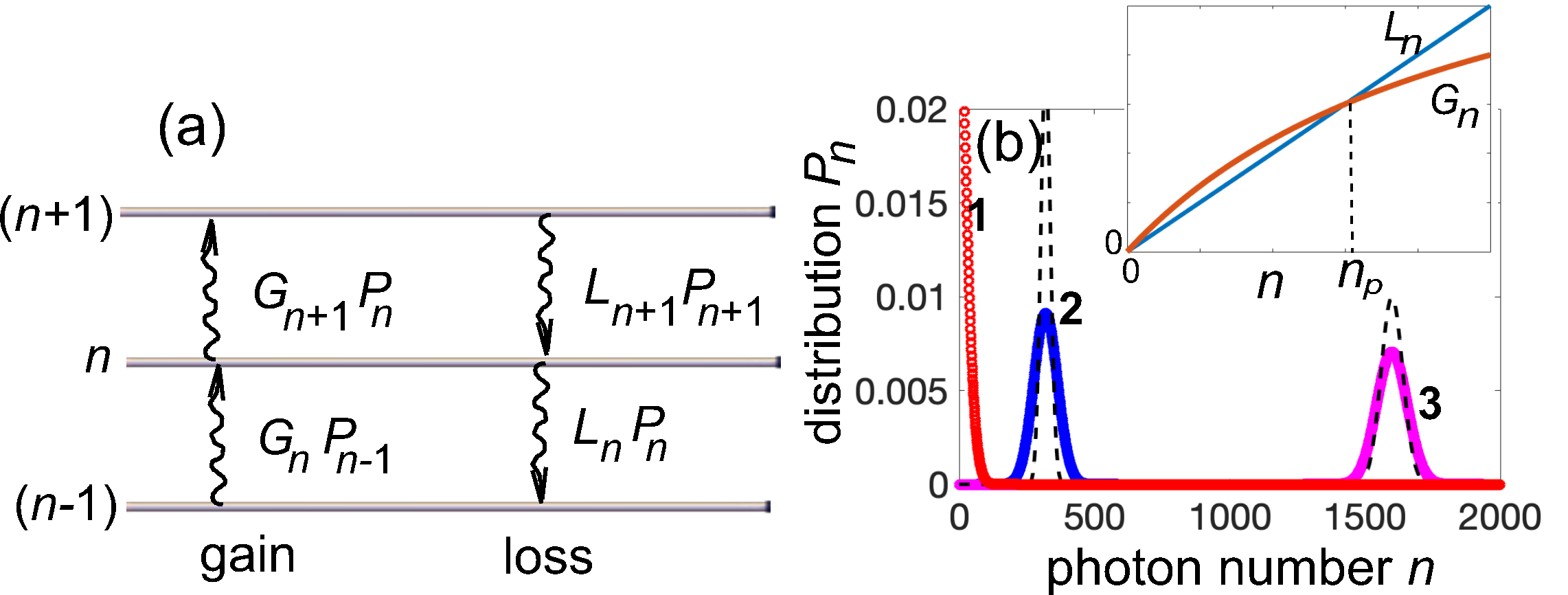}
   \caption{ \small (a) Flow of probability due to stimulated emission and damping entering in the photon master equation (1). (b) Behavior of the stationary photon distribution $P_n^{(S)}$ for $\kappa=1$, $n_s=1600$ and for a few increasing values of the gain parameter $G$.  Curve 1: $G=0.98$; curve 2: $G=1.2$; curve 3: $G=2$. The dashed curves, partly overlapped with curves 2 and 3, show as a reference the Poisson distributions with the same mean photon number than distributions 2 and 3. A clear phase transition, from nearly exponential (thermal light) to nearly Poissonian (coherent light) distributions is observed as the gain parameter $G$ overcomes the loss rate $\kappa$. The inset in (b) schematically shows the behavior of the gain term, $G_n$,  and loss term, $L_n$, versus $n$ for a laser above threshold. The peak value of the distribution $P_n^{(S)}$ is obtained at the crossing point $n=n_p$ of the two curves, which is also very close to the mean photon number $\bar{n}$.}
\end{figure}
where the photon creation (birth) and annihilation (death) rates $G_n$ and $L_n$ are given by \cite{R1,R4}
\begin{equation}
G_n= n \frac{G}{1+n/n_s} \; ,\;\; L_n= n \kappa. 
\end{equation}
In the above equation, $\kappa$ is the photon decay rate in the passive optical cavity, $G$ is the linear gain parameter and $n_s$ is the saturation photon number. Its inverse, $(1/n_s)$, equals the spontaneous emission factor, i.e. the fraction of the  total spontaneous emission that is directed into the lasing mode, which in a typical laser system is a very small number ($1/n_s \sim 10^{-3}-10^{-9}$ \cite{R5}). The stationary distribution $P_n^{(s)}$ of Eq.(1) is readily obtained from the detailed balance condition (see e.g. \cite{Carmichael}), 
\begin{equation}
 L_nP_{n}^{(S)}=G_n P_{n-1}^{(s)}, \nonumber
 \end{equation}
 and is given by $P_n^{(S)}=P_{0}^{(S)} \prod_{l=1}^{n}(G_l/L_l)$, i.e. by the displaced Poisson distribution
\begin{equation}
P_n^{(S)}=P_0^{(S)} \left( \frac{G n_s} {\kappa} \right)^n \frac{n_s!}{(n+n_s)!}
\end{equation}
where $P_0^{(S)}$ is determined from the normalization condition $\sum_{n=0}^{\infty} P_n^{(S)}=1$. 
As is well known, for $n_s \gg 1$ the photon distribution (3) undergoes a phase transition at $G/ \kappa=1$, i.e. at laser threshold, with $P_{n}^{(S)}$ being well approximated by a thermal distribution (chaotic light) with a mean photon number $\bar{n}=G/(\kappa-G)$ below threshold ($G< \kappa$), and by a  nearly-Poisson distribution (coherent light) with mean photon number $\bar {n }=n_s(G/ \kappa-1)$ and variance $\bar{n}+n_s$ above threshold ($G> \kappa$) \cite{R3,R4,R5}; see Fig.1(b).\\ 
  \begin{figure}[h]
\centering
    \includegraphics[width=0.48\textwidth]{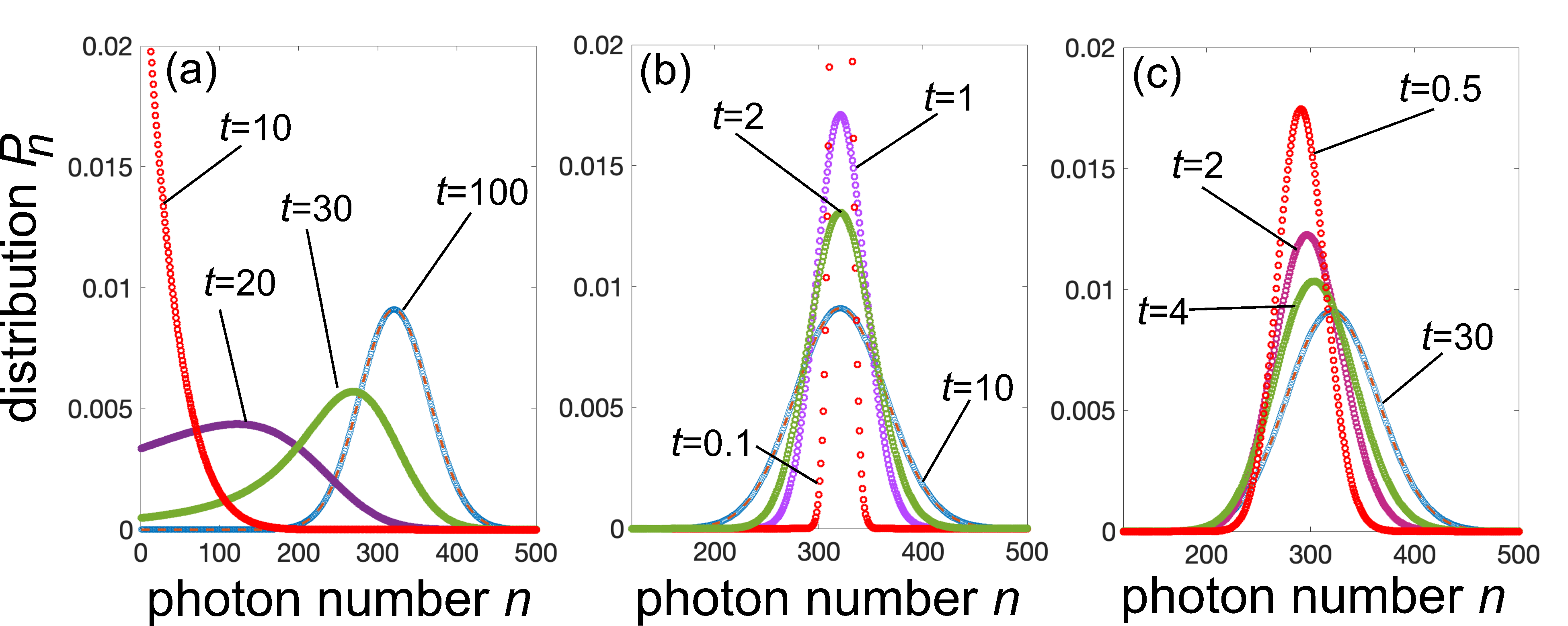}
   \caption{\small Numerically-computed behavior of the relaxation dynamics of the photon distribution probability $P_n(t)$ toward equilibrium in a laser above threshold for parameter values $\kappa=1$, $G=1.2$ and $n_s=1600$ and for a few different initial distributions: (a) vacuum state $P_n(0)=\delta_{n,0}$; (b) Fock state $P_n(0)=\delta_{n,\bar{n}}$ with number $\bar{n}=n_s(G/\kappa-1)=320$; (c) Poisson distribution $P_n(0)=(\bar{n}_1^n/ n!) \exp(-\bar{n}_1)$ with mean photon number $\bar{n}_1=0.9 \times \bar{n}$. The dashed curve in the panels depicts the stationary photon distribution.}
\end{figure}
For a given initial distribution $P_n(0)$ of photons in the cavity mode, the  relaxation towards the  stationary distribution $P_n^{(S)}$ can be determined numerically  solving Eq.(1) \cite{R12}. As an illustrative example, Fig.2 depicts the transient relaxation dynamics in a laser above threshold ($G/ \kappa=1.2)$ for three distinct initial photon distributions: the vacuum state, $P_n(0)=\delta_{n,0}$ [panel(a)]; a single Fock state with a photon number  $n=\bar{n}$, $P_n(0)=\delta_{n,\bar{n}}$ [panel (b)]; and a Poisson distribution with mean photon number $\bar{n}_1=0.9 \bar{n}$ [panel (c)].
Notably, the relaxation to the stationary distribution takes significantly longer in case (a), as there are no photons initially in the cavity. Interestingly, however, the relaxation process is faster in case (b) than in case (c), even though the initial photon distribution in (b) is farther from equilibrium than the distribution in (c). This counterintuitive behavior, which highlights the Mpemba effect, will be explored in detail later.\\
The relaxation dynamics is established by the spectral properties of the Markov transition matrix $M$ entering in the Scully-Lamb equations (1), which is a tridiagonal (Jacobi) semi-infinite matrix, and of its Hermitian adjoint $M^{\dag}$. We indicate by $\lambda_{\alpha}$ ($\alpha=0,1,2,3,...$) the eigenvalues of $M$, by $\psi_n^{(\alpha)}$ the corresponding right eigenfunctions (or just eigenfunctions), and by $\phi_n^{(\alpha)}$ the left eigenfunctions of $M$, i.e. 
$M \psi_n^{(\alpha)}= \lambda_{\alpha} \psi_n^{(\alpha)}$ and $M^{\dag} \phi_n^{(\alpha)}= \lambda_{\alpha}  \phi_n^{(\alpha)}$.  The eigenfunctions are normalized such that $\sum_n \phi_n^{(\alpha)} \psi_n^{(\beta)}= \delta_{\alpha, \beta}$. The eigenvalues $\lambda_{\alpha}$ are real and ordered such that 
$...<\lambda_3< \lambda_2 < \lambda_1<\lambda_0=0$. The right eigenfunction corresponding to the  zero eigenvalue $\lambda_0=0$ is the stationary distribution, i.e. $\psi_n^{(0)}=P_n^{(S)}$, with the corresponding left eigenfunction $\phi_n^{(0)}=1$. When the laser is below threshold and the gain saturation term in the equations is negligible, i.e. by letting $G_n \simeq Gn$, the problem is integrable using the characteristic function and moment methods \cite{R37}. Here we are interested in the laser operation above threshold, where gain saturation is non-negligible and light emission is coherent. Unfortunately, in this regime the birth-death process described by the master equation (1) is not integrable using the characteristic function method \cite{R38}, and one has to resort to full numerical analysis or to approximate techniques. In the large $n_s$ limit and for the laser well above threshold, an approximate expression of the low-order eigenvalues $ \lambda_{\alpha}$ and corresponding right/left eigenfunctions can be obtained by an asymptotic analysis, which is presented in Sec.1 of  the Supplemental document. Namely, one has
\begin{eqnarray}
\lambda_{\alpha} & \simeq & -\alpha(\kappa-g) \\
\phi_n^{(\alpha)} & \simeq & \frac{1}{ \alpha ! 2 ^{\alpha} }H_{\alpha} \left(  \frac{n-n_p}{\sqrt{2(n_p+n_s)}}  \right) \\
\psi_n^{(\alpha)} & \simeq & P_n^{(S)}  H_{\alpha} \left(  \frac{n-n_p}{\sqrt{2(n_p+n_s)}}  \right) 
\end{eqnarray}
 where $H_{\alpha}(x)$ is the Hermite polynomial of order $\alpha$ ($\alpha=0,1,2,3,...$). In the above equations, $n_p=n_s(G/ \kappa-1) \simeq \bar{n}$ is the photon number at which the stationary distribution has its peak, which is obtained by letting $L_{n_p}=G_{n_p}$ and is almost equal to the mean photon number $\bar{n}$; and  
 \[ g=\left( \frac{dG}{dn} \right)_{n_p}=\frac{\kappa^2}{G} \]
 is the differential gain at $n=n_p$, which is clearly smaller than the loss rate $\kappa$ [see the inset in Fig.1(b)].
\par
 \begin{figure}[h]
\centering
    \includegraphics[width=0.43\textwidth]{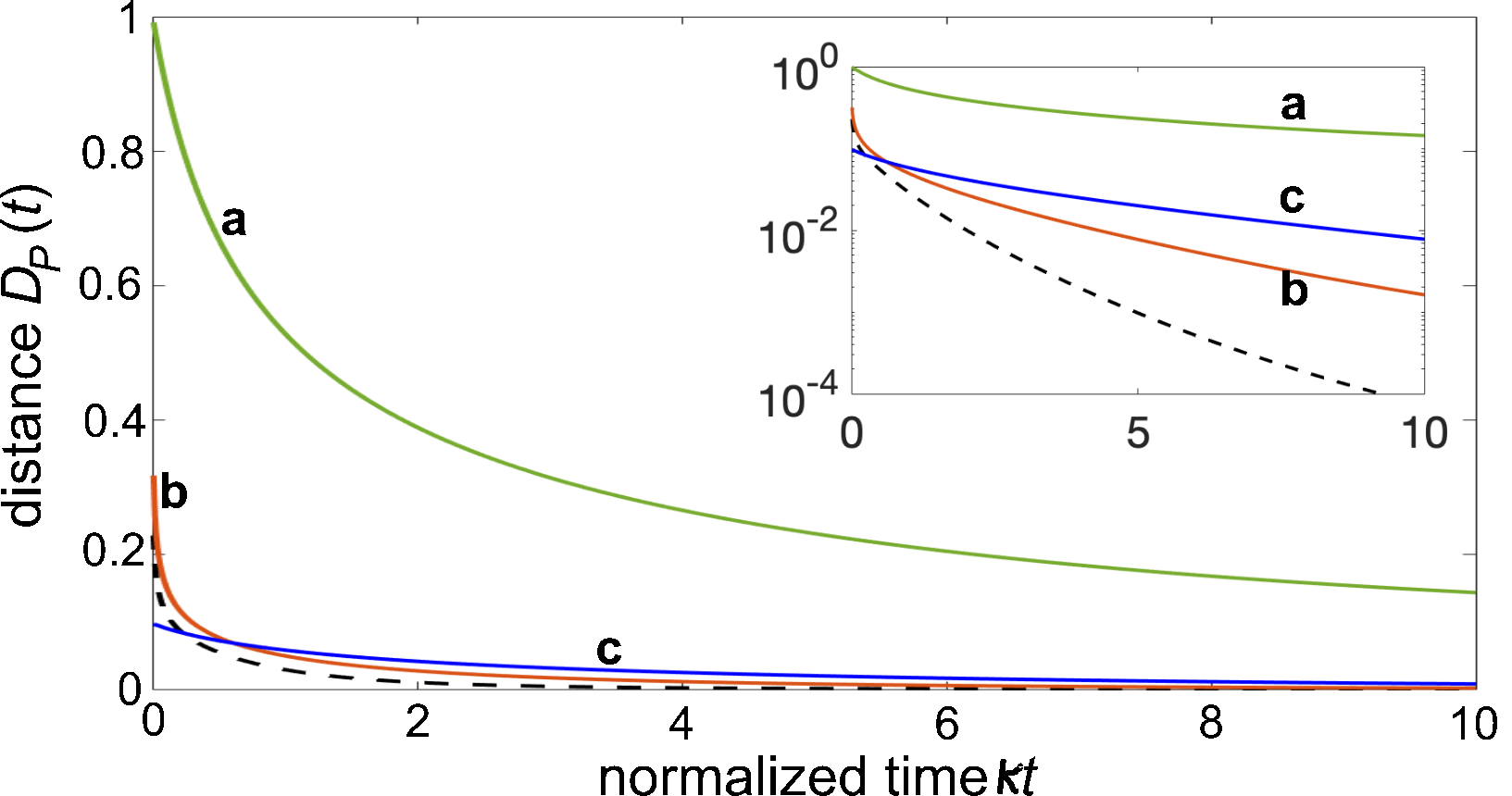}
   \caption{\small Temporal evolution of the distance $D_P(t)$from the equilibrium photon distribution in a laser above threshold corresponding to the three different initial states a, b and c of Fig.2. The inset depicts the temporal dynamics using a log scale in the vertical axis. The dashed curve, corresponding to a super-accelerated relaxation process, depicts the evolution of the distance $D_P(t)$ for the initial photon distribution $P_n(0)=(1/2)(\delta_{n,\bar{n}-\sqrt{\bar{n}+n_s}}+\delta_{n,\bar{n}+\sqrt{\bar{n}+n_s}})  $, i.e. for a mixed two Fock state with the same mean ($\bar{n}=n_p=320$) and variance ($\bar{n}+n_s=1920$) of the equilibrium distribution. }
\end{figure}
{\it Mpemba effect.}
Let us consider an initial photon distribution $P_n(0)$ with $P_n(0) \rightarrow 0$ fast enough as $n \rightarrow \infty$, such that all moments of the distribution, $ \sum_{n=0}^{\infty}n^lP_n(0)$ ($l=1,2,3,...$), are finite. In this case the solution to the laser master equation (1) can be written as a superposition of the right eigenfunctions of $M$, i.e.
\begin{equation}
P_n(t)=\sum_{\alpha=0}^{\infty} C_{\alpha} \psi_n^{(\alpha)} \exp(\lambda_{\alpha}t)=P_n^{(S)}+\sum_{\alpha=1}^{\infty} C_{\alpha} \psi_n^{(\alpha)} \exp(\lambda_{\alpha}t)
\end{equation}
where the spectral amplitudes $C_{\alpha}$ are given in terms of scalar products involving the left eigenfunctions of $M$,
\begin{equation}
C_{\alpha}=\sum_{n=0}^{\infty} P_n(0) \phi_n^{(\alpha)}
\end{equation}
with $C_0=\sum_n P_n(0)=1$. Equation (7) shows that, since $\lambda_{\alpha}=- \alpha (\kappa-g)<0$ for $\alpha>0$, as $t \rightarrow \infty$ the distribution $P_n(t)$ approaches the equilibrium distribution $P_n^{(S)}$. The relaxation time toward equilibrium is determined by the slowest decaying mode entering in the power series of Eq.(7). For the low-order eigenfunctions, we can use the approximate form Eq.(5) for the left eigenfunctions, so that the spectral amplitude $C_1$ turns out to be approximately equal to
\begin{equation}
C_1\simeq \sum_{n=0}^{\infty} (n-\bar{n}) P_n(0).
\end{equation}
 For a rather generic initial state $P_n(0)$ such that $C_1 \neq 0$, the relaxation rate is given by $|\lambda_1|= \kappa-g$. However, for initial states such that $C_1=0$, i.e. when the mean photon number of the distribution $P_n(0)$ is precisely $\bar{n} \simeq n_p$,  the relaxation is faster, with a rate no smaller than $| \lambda_2|=2(\kappa-g)$. 
This result indicates that the photon relaxation dynamics toward the nearly-Poisson distribution can be accelerated for certain initial states and the Mpemba effect can emerge, as discussed in several previous works (see e.g. \cite{R24}). To quantify the relaxation dynamics, one should introduce a measure of the distance $D_P(t)$ of the transient photon distribution $P_n(t)$  from the equilibrium distribution  $P_n^{(S)}$. The definition of the distance $D_{P}(t)$
 is not  unique, and different measures have been considered in previous works, including the Hilbert-Schmidt distance, the trace distance, the Kullback-Leibler divergence and the entanglement asymmetry, to mention a few. Such a diversity in definitions of the distance could raise some ambiguities, i.e.  whether different measures yield consistent results, an issue that is still debated and a subject of recent studies (see e.g. \cite{R34b}).
 Here we adopt he Hilbert-Schmidt distance, defined as in Ref.\cite{R24}
  \begin{eqnarray}
  D_{P}(t)=\sqrt{\sum_{n=0}^{\infty} \left( P_n(t) -P_n^{(S)} \right)^2}.
  \end{eqnarray}
 which is basically the $l^2$ norm of the vector $P_n(t)-P_n^{(S)}$. Note that, if the relaxation rate of $P_n(t)$ toward the equilibrium distribution is $ |\lambda_{\alpha}|$ [the leading non-vanishing term in the series on the right hand side of  Eq.(7)], then asymptotically one has $D_P(t) \rightarrow 0$ exponentially at the same rate $ |\lambda_{\alpha}|$. 
    The Mpemba effect arises whenever, for two assigned initial photon distributions $P_n^{(I)}(0)$ and  $P_n^{(II)}(0)$ in the cavity mode such that  $D_{P^{(II)}} (0)>D_{P^{(I)}} (0)$, asymptotically for $t \rightarrow \infty$ one has $D_{P^{(II)}} (t)<D_{P^{(I)}} (t)$. In other words, the initially state further from the equilibrium relaxes faster. An example of the Mpemba effect is shown in Fig.3, which depicts the numerically-computed behavior of the distance $D_P(t)$ corresponding to the three relaxation processes (a), (b) and (c) of Fig.2. As anticipated, the initial Fock state distribution at the photon number $n=\bar{n}$  [Fig.2(b)] is farther from the equilibrium than the Poisson distribution with mean number $\bar{n}_1=0.9 \times \bar{n}$ [Fig.2(c)]. However, it relaxes faster because the mean photon number of distribution (b) is exactly $\bar{n}$, whereas distribution (c) does not meet this condition. This result is independent of the ratio $G/ \kappa$ and is observed even for a laser operating well above threshold, such as when $G/ \kappa=2$, with the main difference being a faster relaxation toward the equilibrium photon distribution (see Fig. S2 in the Supplemental document).\\  
 Finally, we briefly mention that, if the initial non-equilibrium photon distribution, besides of having a  mean photon number equal to $\bar{n}$, has also a variance equal to the one of $P_n^{(S)}$, i.e. $\sum_{n=0}^{\infty} (n-\bar{n})^2P_n(0)=\bar{n}+n_s$,  in the power series expansion (7) both spectral coefficients $C_1$ and $C_2$ vanish, as it readily follows from Eq.(8) and using for the left eigenfunctions $\phi_n^{(\alpha)}$ the approximate form given in terms of Hermite polynomials [Eq.(5)]. In this case the relaxation toward equilibrium is even faster, i.e. the relaxation process is super-accelerated, a phenomenon similar to the one observed for thermal light \cite{R37}. An example of super-acceleration in the equilibration dynamics is shown by the dashed curve in Fig.3, which depicts the relaxation dynamics of the initial mixed two Fock state 
\begin{equation}
P_n(0)=(1/2) \left( \delta_{n,\bar{n}-\sqrt{\bar{n}+n_s}}+\delta_{n,\bar{n}+\sqrt{\bar{n}+n_s}}   \right).
\end{equation}
This photon distribution displays the same mean and variance than the equilibrium distribution, and thus it relaxes faster than all other initial distributions, including distribution (b).\\
As briefly discussed in Sec.2 of the Supplemental document, the experimental demonstration of the laser Mpemba effect could be  realized by quench dynamics in gain or $Q$-switched class-A lasers \cite{R5bis}, such as the He-Ne laser, where different out-of-equilibrium initial photon distributions can achieved in the cavity.\\
\\
{\em Conclusion.}
This work investigated the emergence of the Mpemba effect in the quantum dynamics of laser light above threshold, using the Scully-Lamb model as a framework. By analyzing the relaxation of photon number statistics under various initial conditions, we uncovered the counterintuitive phenomenon of faster equilibration from states farther from equilibrium. Our findings enrich our understanding of non-equilibrium phenomena in quantum optics and open new avenues for experimental and theoretical exploration of relaxation dynamics in open quantum systems.  Furthermore, the insights gained here may have broader implications for other fields, including Bose-Einstein condensates and atom lasers.\\
\\

\noindent
{\bf Disclosures}. The author declares no conflicts of interest.\\
\\
{\bf Data availability}. No data were generated or analyzed in the presented research.\\
\\
{\bf Funding}. Agencia Estatal de Investigacion (MDM-2017-0711).\\
\\
{\bf Supplemental document}. See Supplement 1 for supporting content.\\

\newpage


 {\bf References with full titles}\\
 \\
 \noindent
 \small
1. M.O. Scully and W.E. Lamb, Jr., Quantum Theory of an Optical Maser. I. General Theory,
Phys. Rev. {\bf 159}, 208 (1967).\\
2.  M. Lax and W.H. Louisell, Noise. XII. Density-Operator Treairiient and Population Fluctuations, Phys. Rev. {\bf 185}, 568 (1969).\\
3. M.O. Scully and M.S. Zubairy, {\em Quantum Optics} (Cambridge University Press, 1997).\\
4. D.F. Walls and G.J. Milburn, {\em Quantum Optics} (Springer, 2008).\\
5. R. Loudon, {\em The Quantum Theory of Light} (Oxford University Press, 2000).\\
6. H. J. Carmichael, {\em Statistical Methods in Quantum Optics 1} (Springer, Berlin, 2002), Chap. 7.\\
7. R. Graham and H. Haken, Laserlight-- first example of a second-order phase transition far away from thermal equilibrium, Z. Phys. {\bf 237}, 31 (1970).\\
8. V. Degiorgio and M. O. Scully, Analogy between the laser threshold region and a second-order phase transition, Phys. Rev. A {\bf 2}, 1170 (1970).\\
9. F.T. Arecchi and V. Degiorgio, Statistical Properties of Laser Radiation During a Transient Buildup, Phys. Rev. A {\bf 3}, 1108 (1971).\\
10. F. Minganti,  I.I. Arkhipov, A. Miranowicz, and F. Nori, Liouvillian spectral collapse in the Scully-Lamb laser model, Phys. Rev. Res. {\bf 3}, 043197 (2021).\\
11. A.M. Yacomotti, Z. Denis, A. Biella, and C. Ciuti, Quantum Density Matrix Theory for a Laser Without Adiabatic Elimination of the Population Inversion: Transition to Lasing in the Class-B Limit,
Laser \& Photon. Rev. {\bf 17}, 2200377 (2023).\\
12. M.O. Scully, Condensation of N Bosons and the Laser Phase Transition Analogy, Phys. Rev. Lett. {\bf 82}, 3927 (1999).\\
13. V. V. Kocharovsky, M.O. Scully, S.-Y.. Zhu, and M.S. Zubairy, Condensation of N bosons. II. Nonequilibrium analysis of an ideal Bose gas and the laser phase-transition analogy,
Phys. Rev. A {\bf 61}, 023609 (2000).\\
14. V.V. Kocharovsky, V.V. Kocharovsky, M. Holthaus, C.H.
Raymond Ooi, A. Svidzinsky, W. Ketterle, and M.O. Scully, 
Fluctuations in ideal and interacting Bose-Einstein condensates: From the laser phase transition analogy to squeezed states and Bogoliubov quasiparticles, 
Adv. At. Mol. Opt. Phys. {\bf 5}3, 291 (2006).\\
15. Z. Zhang, G.S. Agarwal, and M.O. Scully,
Quantum Fluctuations in the Fr\"ohlich Condensate of Molecular Vibrations Driven Far From Equilibrium, Phys. Rev. Lett. {\bf 122}, 158101 (2019).\\
16. M.O. Scully, S. Fulling, D.M. Lee, and A.A. Svidzinsky,
Quantum optics approach to radiation from atoms falling into a black hole,
Proc.Nat. Ac. Sci. PNAS {\bf 115}, 8131 (2018).\\
17. M.O. Scully, A. Svidzinsky, and W. Unruh, On Bose-Einstein Condensation and Unruh-Hawking
Radiation from a Quantum Optical Perspective, J. Low Temp. Phys. {\bf 208}, 160 (2020).\\ 
18. A.S. Novozhilov, G.P. Karev, and E.V. Koonin,
Biological applications of the theory of birth-and-death processes, Brief. Bioinform. {\bf 7}, 70 (2006).\\
19. M. Sargent III, M.O. Scully, and W.E. Lamb Jr., Buildup of laser oscillations from quantum noise, Appl. Opt. {\bf 9}, 2423 (1970).\\
20. P. Mandel, Nonstationary behaviour of a quantum monomode laser, Physica {\bf 63}, 553 (1973).\\
21. D. Walgraef, Quantum statistics of a monomode-laser model, Physica {\bf 72}, 578 (1974).\\
22. J. Fiutak and J. Mizerski, Transient behaviour of laser, Z. Phys. B {\bf 39}, 347 (1980). \\
23. M.-O. Mewes, M.R. Andrews, D.M. Kurn, D.S. Durfee, C.G. Townsend, and W. Ketterle,
Output Coupler for Bose-Einstein Condensed Atoms, Phys. Rev. Lett. {\bf 78}, 582 (1997).\\
24. A. \"{O}ttl, S. Ritter, M. K\"ohl, and T. Esslinger,
Correlations and Counting Statistics of an Atom Laser, Phys. Rev. Lett. {\bf 95}, 090404 (2005).\\
25. E. B. Mpemba and D. G. Osborne, Cool?,
Phys. Educ.\textbf{4}, 172 (1969).\\
26.  M. Jeng, The Mpemba effect: when can hot water freeze faster than cold?,
 Am. J. Phys. \textbf{74}, 514 (2006).\\
27.  A. Lasanta, F. Vega Reyes, A. Prados, and A. Santos, When the Hotter Cools More Quickly: Mpemba Effect in Granular Fluids,
 Phys. Rev. Lett. \textbf{119}, 148001  (2017).\\
28. Z. Lu and O. Raz, Non-equilibrium thermodynamics of the Markovian Mpemba effect and its inverse,Proc. Natl. Acad. Sci. U.S.A.
       \textbf{114}, 5083  (2017).\\
29. A. Nava and M. Fabrizio, Lindblad dissipative dynamics in the presence of phase coexistence, Phys. Rev. B  \textbf{100}, 125102  (2019). \\
30. S.K. Manikandan, Equidistant quenches in few-level quantum systems,
Phys. Rev. Res.  \textbf{3}, 043108   (2021).\\
31. F. Carollo, A. Lassant, and I. Lesanovsky, Exponentially Accelerated Approach to Stationarity in Markovian Open Quantum Systems through the Mpemba Effect,
Phys. Rev. Lett.  \textbf{127},  060401    (2021).\\
32. Y.-L. Zhou, X.-D. Yu, C.-W. Wu,  X.-Q. Li, J. Zhang, W. Li, and P.X. Chen, Accelerating relaxation through Liouvillian exceptional point,
Phys. Rev. Res.  \textbf{5}, 043036 (2023).\\
33. A.K. Chatterjee, S. Takada, and H. Hayakawa, Quantum Mpemba Effect in a Quantum Dot with Reservoirs,
Phys. Rev. Lett.  \textbf{131}, 080402  (2023). \\
34. F. Ares, S. Murciano, and P. Calabrese, Entanglement asymmetry as a probe of symmetry breaking,
Nat. Commun. \textbf{14}, 2036  (2023).\\
35. L. Kh Joshi, J. Franke, A. Rath, F. Ares, S. Murciano, F. Kranzl, R. Blatt, P. Zoller, B. Vermersch, P. Calabrese, C.F. Roos, and M.K. Joshi, Observing the Quantum Mpemba Effect in Quantum Simulations,
Phys. Rev. Lett. \textbf{133}, 010402   (2024).\\
36. C. Rylands, K. Klobas, F. Ares, P. Calabrese, S. Murciano, and B. Bertini. Microscopic origin of the quantum Mpemba effect in integrable systems,
Phys. Rev. Lett. \textbf{133}, 010401 (2024).\\
37. S.A. Shapira, Y. Shapira, J. Markov, G. Teza, N. Akerman, O. Raz, and R. Ozeri,
Inverse Mpemba Effect Demonstrated on a Single Trapped Ion Qubit,
Phys. Rev. Lett. \textbf{133}, 010403 (2024).\\
38.  M. Moroder, O. Culhane, K. Zawadzki, and J. Goold,
Thermodynamics of the Quantum Mpemba Effect,
Phys. Rev. Lett. \textbf{133}, 140404 (2024).\\
39. S. Liu, H.-K. Zhang, S. Yin, and S.-X. Zhang, Symmetry Restoration and Quantum Mpemba Effect in Symmetric Random Circuits,
Phys. Rev. Lett. \textbf{133}, 140405 (2024).\\
40. A. Nava and R. Egger, Mpemba Effects in Open Nonequilibrium Quantum Systems,
Phys. Rev. Lett. \textbf{133}, 136302 (2024).\\
41. J. Zhang, G. Xia, C.-W. Wu, T. Chen, Q. Zhang, Y. Xie, W.-B. Su, W. Wu, C.-W. Qiu, P. xing Chen, W. Li, H. Jing, and Y.-L. Zhou,
Observation of quantum strong Mpemba effect, arXiv:240115951 (2024); to appear in Nat. Commun.\\
42. D. Qian, H. Wang, and J. Wang,
Intrinsic Quantum Mpemba Effect in Markovian Systems and Quantum Circuits, arXiv:2411.18417v1 (2024).\\
43. S. Longhi, Photonic Mpemba effect, Opt. Lett. \textbf{49}, 5188 (2024).\\ 
44. S. Longhi, Bosonic Mpemba effect with non-classical states of light,
APL Quantum {\bf 1},  046110 (2024).\\ 
45. S. Longhi, Mpemba effect and super-accelerated thermalization in the damped quantum harmonic oscillator, arXiv:2411.09589 (2024).\\
46. P.B. Acosta-Humanez, J.A. Capitan, and J.J. Morales-Ruiz, Integrability of Stochastic Birth-Death processes via Differential Galois Theory, Math. Model. Nat. Phenom. {\bf 15}, 70 (2020).\\

\end{document}